\documentclass[twocolumn,floats,showpacs,pra]{revtex4}
\usepackage[above]{placeins}
\usepackage{amsmath}
\usepackage{graphicx}

\begin{document}

\title{New insight into the Hall effect}
\author{X. Q. Huang}
\email{xqhuang@netra.nju.edu.cn}
\affiliation{Department of Physics and National Laboratory of Solid State Microstructure,
Nanjing University, Nanjing 210093, China}
\date{\today}

\begin{abstract}
In this paper, we develop a unified theory for describing Hall effect in
various electronic systems based on a pure electron picture (without the
hole concept). We argue that the Hall effect is the magnetic field induced
symmetry breaking of the charge carrier's spatial distribution. Due to the
interaction of the charge carriers and the ion lattice, there are two
possible symmetry breaking mechanisms which cause different signs of Hall
coefficient in a Hall material. The scenario provides an explicit
explanation of the sign different of the Hall coefficient in the $N$-type
and $P$-type semiconductors, the sign reversal induced by both temperature
and magnetic field in different materials, and the integer and fractional
quantum Hall effect (QHE) in two-dimensional electron gas (2DEG) of
GaAs/AlGaN heterostructures.
\end{abstract}

\maketitle

\section{Introduction}

Since the discovery of the Hall effect over a century ago \cite{Hall}, much
effort have been made to elucidate this phenomenon both experimentally and
theoretically. In the experiment, a number of Hall effect-related phenomena
have been uncovered. Of these, the most remarkable achievements are the
observation of the integer quantum Hall effect (IQHE) \cite{Klitzing} and
the fractional quantum Hall effect (FQHE) \cite{Tsui} in the two-dimensional
electron systems. In the theory, different approaches have been developed
and applied to the study of these peculiar experimental results, however,
the physical interpretation of the Hall effects is made difficult even in
relatively simple metals. It is well known that in many cases the simple
semiclassical expression of the Hall coefficient $R_{H}=1/qn$ does not work,
where $n$ is the density of mobile charges and $q$ is the charge of the
charge carriers. In strongly correlated systems the Hall effect is even more
difficult to interpret because more factors can have a large influence on
the Hall resistivity. As a result, some unexpected reversals of the sign
(the so-called anomalous Hall effect) of the Hall coefficient (or voltage)
have been reported in these systems.

Despite many works have been put into this field, the mechanisms which lead
to the Hall effects are not completely understood. In fact, the theoretical
investigations are almost hampered by the complexity of the system. In this
paper, we would like to address that the theoretical difficulties
encountered in the Hall effect do not arise from the complexity of the
studied systems, but from a fatal misunderstanding of the most fundamental
mechanism that rules the Hall effects. It will be shown clearly that a pure
electron picture (without the hole concept) can provide a unified
explanation of the Hall effects for different electronic systems, for
example, the sign different of the Hall coefficient in the $N$-type and $P$%
-type semiconductors, the sign reversal in different materials, and the
appearance of the plateaux in the Hall conductivity in two-dimensional
electron systems at low temperatures and in strong magnetic fields.

\begin{figure}[ph]
\begin{center}
\resizebox{1\columnwidth}{!}{
\includegraphics{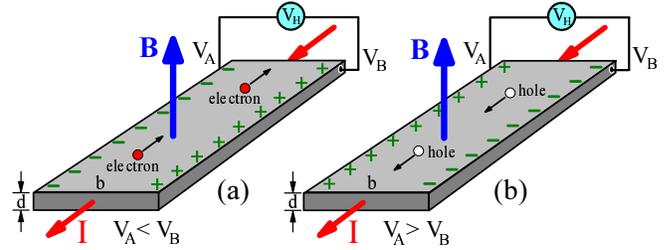}}
\end{center}
\caption{The schematic plot of Hall effect with different kinds of charge
carriers. (a) Negative charge carriers: electrons, and (b) positive charge
carriers: holes. }
\label{fig1}
\end{figure}

\section{An old Look for Hall effect}

According to the traditional viewpoint of the Hall effect, the basic
physical reason underlying the Hall effect is the Lorentz force. As shown in
Fig. \ref{fig1}, when an electric current flows through a conductor in a
direction perpendicular to an applied magnetic field which exerts a
transverse force on the moving charge carriers (the electric current), this
force tends to push the moving charges to one side of the conductor and a
Hall voltage ($V_{H}=V_{A}-V_{B}$) builds up. Note that the Fig. \ref{fig1}%
(a) is for $N$-type materials with the negative carriers (electrons) while
Fig. \ref{fig1}(b) for $P$-type materials with the positive charge carriers
(holes). Normally, the $N$-type materials have a negative Hall voltage,
while the $P$-type materials have a positive Hall voltage.

\begin{figure}[tbp]
\begin{center}
\resizebox{1\columnwidth}{!}{
\includegraphics{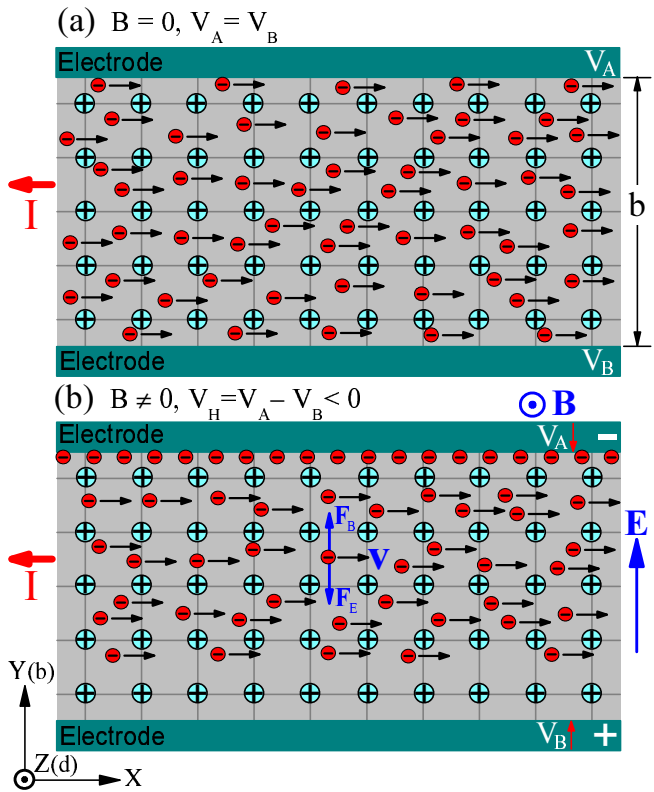}}
\end{center}
\caption{A qualitative interpretation of Hall effect for a simple metal. (a)
Without an external magnetic field, (b) under an external magnetic field. }
\label{fig2}
\end{figure}

Figure \ref{fig2} illustrates the Hall effect for a simple metal where there
is only one type of charge carrier (electrons). Without an external magnetic
field, there is no Hall effect ($V_{H}=0$) in the sample, as shown in Fig. %
\ref{fig2}(a). When a magnetic field is applied in $Z$-direction, a buildup
of charge at the sides of the conductors (characterized by an static
electric field $\mathbf{E}$ in $Y$-direction) will balance this magnetic
influence, producing a measurable Hall voltage $V_{H}$ which is given by

\begin{equation}
V_{H}=-\frac{1}{n_{e}e}\frac{IB}{d}=R_{H}\frac{IB}{d},  \label{hall1}
\end{equation}%
where $n_{e}$ is the charge carrier density of the carrier electrons, $B$ is
the magnetic flux density, $d$ is the depth of the sample, $e$ is the
electron charge, and the electric current can be expressed in terms of the
drift velocity $I=-n_{e}evbd$. It is easy to find that Eq. (\ref{hall1})
leads to the following expression%
\begin{equation}
n_{e}eR_{H}=-1.  \label{hall2}
\end{equation}

Eq. (\ref{hall2}) implies that the Hall coefficient $R_{H}$ of the simple
metals should always be negative. Table \ref{table1} shows some experimental
values of $1/n_{e}eR_{H}$ for some simple metals. It is not difficult to
find that all the experimental data are inconsistent with the theoretical
prediction. However, probably the most surprising result is that the sign of
$n_{e}eR_{H}$ (or $R_{H}$) is positive for $Li$ and $Na$. We refer this
situation of the Hall effect as \textquotedblleft sign
catastrophe\textquotedblright . To the best of our knowledge, this
\textquotedblleft catastrophe\textquotedblright\ is a rather common
experimental fact that can be found easily in any Hall materials. Many
researchers believe that this problem can be easily overcome by introducing
the concept of the positive \textquotedblleft hole\textquotedblright .
However, our point of view is quite different in this paper. We consider
that one should be cautious when drawing conclusions based on the man-made
concept of \textquotedblleft hole\textquotedblright . In fact, using the
language of `hole' rather than `electron' can obscure the essential physics
of the Hall effect. There is no reason for us to overlook one basic fact
that it is the real electrons, not the artificial holes, carrying the
electric current in the materials. In other words, we need a new look for
Hall effect.

\begin{table}[tbp]
\caption{The experimental values of $1/n_{e}eR_{H}$ for some simple metals.}
\label{table1}$%
\begin{tabular}{ccc}
\hline\hline
Metal element & Number of valence electrons & $1/n_{e}eR_{H}$ \\ \hline
$Li$ & 1 & $0.8$ \\ \hline
$Na$ & 1 & $1.2$ \\ \hline
$Be$ & 2 & $-0.2$ \\ \hline
$Mn$ & 3 & $-0.4$ \\ \hline
$Al$ & 3 & $-0.3$ \\ \hline
$In$ & 3 & $-0.3$ \\ \hline\hline
\end{tabular}%
$%
\end{table}

\begin{figure}[bp]
\begin{center}
\resizebox{1\columnwidth}{!}{
\includegraphics{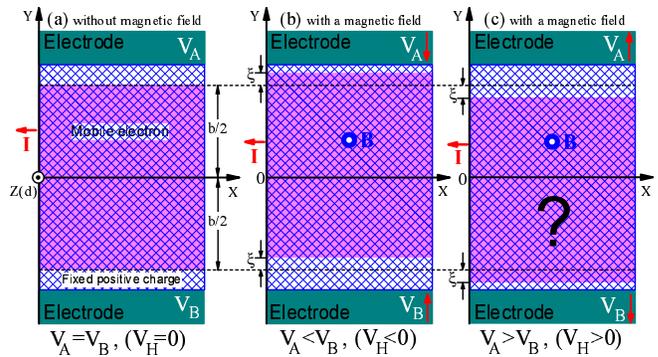}}
\end{center}
\caption{The schematic interpretation of Hall effect in a new angle. (a) A
Hall material contains merely the mobile electrons and the fixed positive
lattice ions, without the external magnetic field, the space distribution of
the mobile electrons has $X$-axis symmetry, (b) normally, with a magnetic
field in $Z$-direction, the mobile electrons are shifted upward by a
interval $\protect\xi $ and cause a negative Hall voltage, (c) we suggest
that the $Z$-direction magnetic field can also induce a abnormal downward
displacement of the mobile electrons and contribute a positive Hall voltage.
}
\label{fig3}
\end{figure}

\section{New look for Hall effect}

Figure \ref{fig3} shows our new idea of Hall effect. For the case without an
external magnetic field, there will be no Hall effect ($V_{H}=V_{A}-V_{B}=0$%
) as shown in Fig. \ref{fig3} (a). When a magnetic field is present in $Z$%
-direction, the mobile electrons have an unitary displacement ($\xi $) along
$+Y$-direction because of the Lorentz force on the electrons, leading to a
negative Hall effect ($V_{H}<0$) as shown in Fig. \ref{fig3} (b). Here we
assume that, under some special circumstance, the mobile electrons can move
in an opposite direction ($-Y$) under the interaction of the magnetic field
and shows a positive Hall effect ($V_{H}>0$) as shown in Fig. \ref{fig3}
(c). A more detailed explanation for such an inexplicable assumption will be
given in the following.

Evidently, as compared to the old picture of Hall effect (see the previous
section), our new idea described in Fig. \ref{fig3} contains four main
innovations:

(1) The new picture does not involve any quasiparticle (for example, the
well-known `hole'), electrons are the only charge carriers in any Hall
materials.

(2) In Fig. \ref{fig3}, the mobile electrons are no longer assigned to be
the \textquotedblleft free electrons\textquotedblright , the interaction
between the mobile electrons and the fixed positive lattice ions should be
taken into account.

(3) We do not think that the magnetic field would cause the accumulation of
the charge carriers at the sides of the Hall materials, in our opinion, the
application of the magnetic field can only result in an overall shift of the
mobile electrons.

(4) The new mechanism is based on the viewpoint of the symmetry breaking,
the magnetic field can induce two different symmetry breaking effects which
contribute to two different Hall effects: the negative Hall effect of Fig. %
\ref{fig3}(b) and the positive Hall effect of Fig. \ref{fig3}(c).

\begin{figure}[tbp]
\begin{center}
\resizebox{1\columnwidth}{!}{
\includegraphics{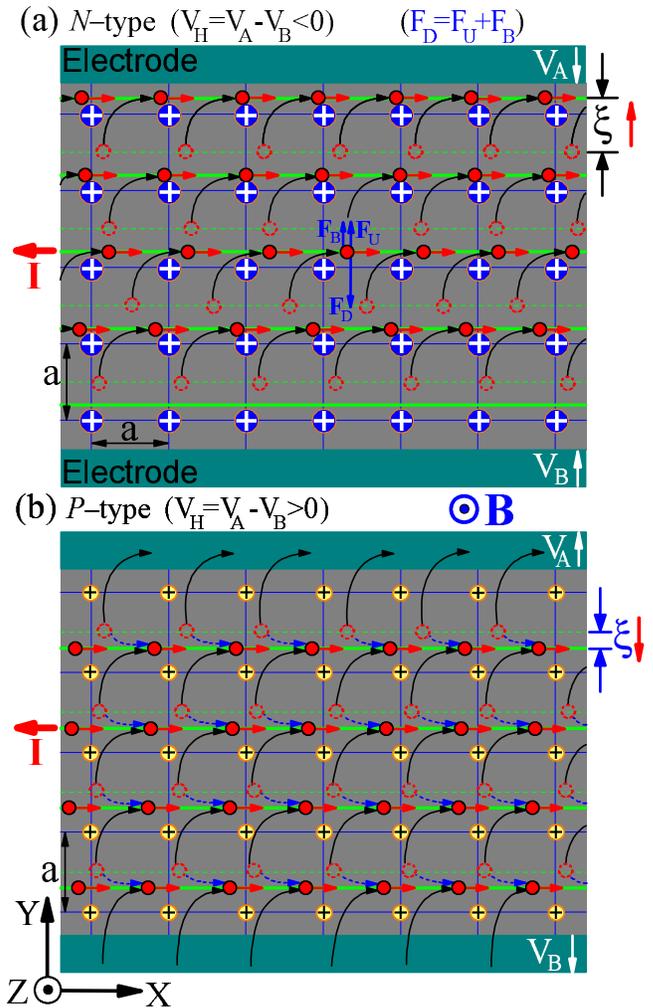}}
\end{center}
\caption{A schematic interpretation of $N$-type \ and $P$-type Hall effects
in a simple metal. (a) A metal with a large number of valence electrons
tends to present a $N$-type Hall effect, because the mobile electrons can be
well confined near the positive ion chains, (b) a metal with a small number
of valence electrons may be a $P$-type of Hall material, some mobile
electrons may enter the electrode due to a relatively weak confinement by
the ion lattice. In this case, the Lorentz force induced upward movement of
the mobile electrons is equivalent to a abnormal downward movement.}
\label{fig4}
\end{figure}

Of course, the key point of the new concept is: As illustrated in Fig. \ref%
{fig3} (c), how can the electrons move in a \textquotedblleft wrong
direction\textquotedblright\ under the magnetic field?

We believe that the sign of Hall coefficient $R_{H}$ is dominated by the
strength of restriction the mobile electrons in a Hall material. When the
mobile electrons are strongly confined inside a material, the corresponding
sample tends to have a negative Hall coefficient, while for a
weak-restricted system, the Hall coefficient most probably has a positive
sign. To illustrate this idea more clearly, I would like to present in Fig. %
\ref{fig4} why a material can exhibit a positive Hall coefficient. Figure %
\ref{fig4}(a) shows a Hall material of a large number of valence electrons,
such as $Mn^{3+}$, $Al^{3+}$, $In^{3+}$ and $Be^{2+}$, which has a negative
sign of the Hall effect ($V_{H}<0$), or the $N$-type Hall effect. As can be
seen from Fig. \ref{fig4}(a), when applied magnetic field, the mobile
electrons will move from the old equilibrium positions (the dash green
lines) to new ones ($F_{D}=F_{U}+F_{B}$) of the solid green lines which are
very close to the positive quasi-one-dimensional ion chains, owing to the
strong electromagnetic interactions between the mobile electrons and the
positive ions. For a small number of valence electrons's Hall material (for
example, $Li^{1+}$, $Na^{1+}$), the influence of an external magnetic field
may be different. As shown in Fig. \ref{fig4}(b), the mobile electrons
nearest to the electrode (A) are most likely to enter into the electrode
directly due to a relatively weak electromagnetic interaction between these
electrons and the ion chain. In this situation, the effect (indicated by the
black solid curved arrows) of the external magnetic field ($+Z$-direction)
on the mobile electrons is equivalent to that (indicated by the blue dash
curved arrows) of a $-Z$-direction magnetic field. Obviously, Fig. \ref{fig4}%
(b) favors the positive sign of the Hall effect ($V_{H}>0$), or the $P$-type
Hall effect.

Based on the new picture of Fig. \ref{fig4}, we now continue to discuss the
dependence of the sign of Hall effect on the properties of the Hall
materials and external factors (for example, the magnetic field and the
temperature).

It's now become very clear that to obtain a positive Hall effect ($P$-type),
the mobile electrons inside the Hall material should be less confined by the
positive lattice ions. By decreasing the number of valence electrons is an
effective way to achieve the positive Hall effect as discussed above. In the
semiconductors, it is well known that the intrinsic silicon may exhibit a $N$%
-type semiconductor and $P$-type behavior through the doping of pentavalent
impurities (such as antimony, arsenic or phosphorous) and trivalent
impurities (such as boron, aluminum or gallium), respectively. From the
viewpoint of Fig. \ref{fig4}, the adding of the pentavalent impurities
greatly enhance the confinement ability of the positive ion lattice on the
mobile electrons, hence the material should have a negative Hall coefficient
($N$-type semiconductor), while the situation for the later case is
different, the trivalent impurities will decrease the confinement ability on
the mobile electrons and lead to a $P$-type Hall behavior.

Besides, there are two important factors (magnetic field and temperature)
which could lead to the $N$-type to $P$-type transition in a Hall material.
We suppose that a material appears to be a $N$-type Hall material under an
external magnetic field $\mathbf{B}$. With the increasing strength of the
magnetic field, some mobile electrons may be free from the restriction of
the positive lattice ions and cause the $N$-type to $P$-type transition in
the material. The same discussion can be used to explain the temperature
induced $N$-type to $P$-type transition, it is quite clear that a higher
temperature would decrease the stability of the ion lattice which in turn
make the mobile electrons less confinement, this fact implies that high
temperature may favor a $P$-type Hall phase. Let us mention that these
discussions can be applied to interpret the observed sign reversal induced
by both temperature and magnetic field in different materials.

\section{Analytical results}

Our scenario of the Hall effect offers a totally new picture of Hall effect
and shows that electric currents in Hall materials are carried only by
moving electrons, not by holes. In other words, it is inappropriate to think
that there are two kinds of charge carriers (electrons and holes) coexist in
materials. In this section, we try to establish the relationships between
Hall voltage $V_{H}$, Hall resistance $R_{xx}$, charge carrier density $%
n_{e} $ and magnetic field $B$ using the new mechanism.

For simplicity, it is assumed that two electrons (denoted by $i$ and $-i$)
are moving along the $X$-direction at velocity $v$ in a positive-charge
background, as shown in Fig. \ref{fig5}. Without the magnetic field, these
two electrons locate in the $Y$-direction at $y_{i}$ and $-y_{i}$,
respectively. If a magnetic field is present in $Z$-direction, the electrons
will be driven $y_{i}+\xi $ and $-y_{i}+\xi $ by the Lorentz force (see Fig. %
\ref{fig5}). As a result, the voltage of the electrode $A$ and $B$ can be
directly given by

\begin{eqnarray}
V_{A}^{i} &=&-\frac{e^{2}}{4\pi \varepsilon _{0}(b/2-y_{i}-\xi )}-\frac{e^{2}%
}{4\pi \varepsilon _{0}(b/2+y_{i}-\xi )},  \label{viA} \\
V_{B}^{i} &=&-\frac{e^{2}}{4\pi \varepsilon _{0}(b/2-y_{i}+\xi )}-\frac{e^{2}%
}{4\pi \varepsilon _{0}(b/2+y_{i}+\xi )}.  \label{viB}
\end{eqnarray}%
Because $\xi \ll b/2$, then Eqs. (\ref{viA}) and (\ref{viB}) lead to the
following expression for the Hall voltage:

\begin{figure}[tbp]
\begin{center}
\resizebox{1\columnwidth}{!}{
\includegraphics{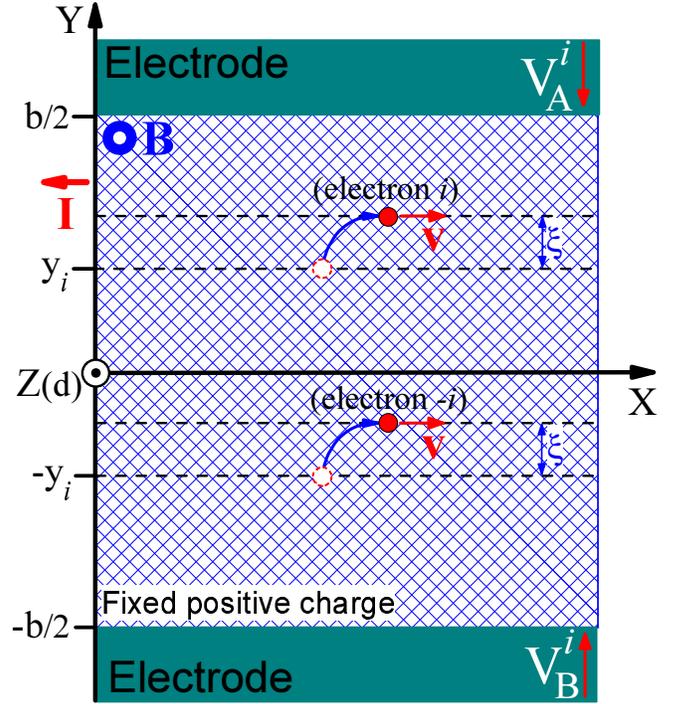}}
\end{center}
\caption{A simply picture of two electrons Hall effect. Under an external
magnetic field, these mobile electrons will move upward with a interval $%
\protect\xi $ due to the Lorentz force and lead to a Hall voltage $%
V_{H}^{i}=V_{A}^{i}-V_{B}^{i}$. }
\label{fig5}
\end{figure}

\begin{eqnarray}
V_{H}^{i} &=&V_{A}^{i}-V_{B}^{i}  \notag \\
&\approx &-\frac{4e^{2}\left( b^{2}+4y_{i}^{2}\right) \xi }{\pi \varepsilon
_{0}\left( b^{2}-4y_{i}^{2}\right) ^{2}}.  \label{vhi}
\end{eqnarray}%
For a real Hall system with a electron density $n_{e}$, the total Hall
voltages is approximately given by%
\begin{equation}
V_{H}=\sum_{i}V_{H}^{i}\propto n_{e}^{\alpha }\frac{\xi }{b^{2}},  \label{vh}
\end{equation}%
where $0<\alpha <1$ is a material related constant. The shift of the mobile
electrons $\xi $ is described by the Lorentz force
\begin{equation}
\xi \propto vB=\frac{n_{e}evbd}{n_{e}ebd}B=-\frac{IB}{n_{e}ebd}.
\label{shift}
\end{equation}%
Inserting Eq. (\ref{shift}) into Eq. (\ref{vh}) one arrives at the Hall
voltage
\begin{equation}
V_{H}=-\kappa \frac{1}{n_{e}^{1-\alpha }eb^{3}}\frac{IB}{d},  \label{hall3}
\end{equation}%
and the Hall resistance
\begin{equation}
R_{xx}=\frac{V_{H}}{I}=-\kappa \frac{B}{n_{e}^{1-\alpha }eb^{3}d},
\label{re}
\end{equation}%
where $\kappa $ is a constant. Although our theoretical analysis above
requires some mathematical approximation, it is easily seen that our result
of Eq. (\ref{hall3}) contains more reasonable terms than the traditional
result of Eq. (\ref{hall1}). Normally, the Hall resistance $R_{xx}$
increases linearly with magnetic field.

\section{Integer and fractional quantum Hall effects}

Undoubtedly, the discoveries of the integer and fractional quantum Hall
effects (QHE) are remarkable achievement in condensed matter physics.
However, the theoretical explanation of these experiments is far beyond
expectations. In fact, a general theoretical understanding of the QHE is
still lacking, many existing theories of QHE have endowed these findings
with some mystery and some apparent contradictory behavior. We are confident
that the new picture of this paper can provide the most simple
straightforward description of the QHE.

\begin{figure}[bp]
\begin{center}
\resizebox{1\columnwidth}{!}{
\includegraphics{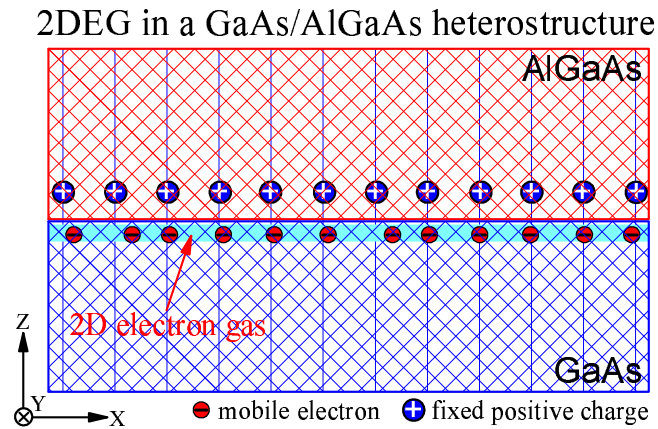}}
\end{center}
\caption{Two dimensional electron gas (2DEG) in a GaAs/GaAlAs
heterojunction. A superlattice of positive fixed charges is formed in GaAlAs
layer. }
\label{fig6}
\end{figure}

\begin{figure}[tp]
\begin{center}
\resizebox{1\columnwidth}{!}{
\includegraphics{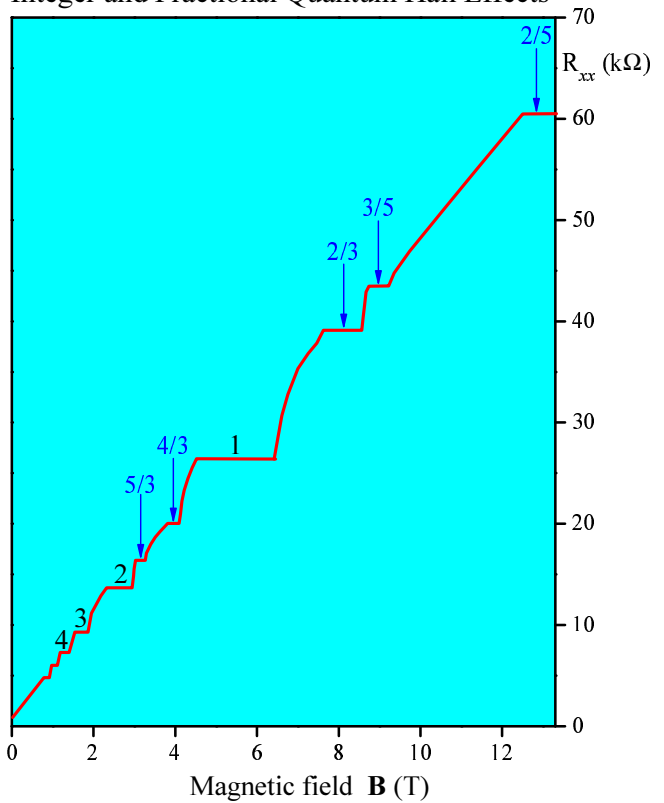}}
\end{center}
\caption{The experimental results of the integer quantum Hall effect and the
fractional quantum Hall effect. Two kind of the Hall resistance plateaux
(integer and fractional) are clearly shown. }
\label{fig7}
\end{figure}

The quantum Hall phenomena can only be observed in the two-dimensional
electron systems (as illustrated in Fig. \ref{fig6}) subjected to extreme
low temperatures and very strong magnetic fields. Figure \ref{fig7} shows
the experimental results of the integer quantum Hall effect (IQHE) and the
fractional quantum Hall effect (FQHE) in a GaAs/GaAlAs heterojunction with a
surface mobile electron density $n_{s}$ about $1.0\times 10^{11}/cm^{2}$.
From the figure it is easy to see that there are two classes of Hall
resistance plateaux, one class is the so-called the integer quantum Hall
effect and the corresponding plateaux can be well described by
\begin{equation}
R_{xx}=\frac{h}{e^{2}N},N=1,2,3,4,5...
\end{equation}%
The other class is referred to as the fractional quantum Hall effect. Here
the Hall resistance plateaux are seen at the so-called fractional filling
factors, especially 1/3, 2/3, 4/3, 5/3, 2/5 and 3/5.

The majority researchers hold that IQHE and FQHE are two absolutely
different types of physical effects which are caused by different physical
reasons, therefore, different theories are needed to explain them. We think
that such a viewpoint is physically inappropriate, or even wrong. We insist
that these two effects share exactly the same physical mechanism.
Furthermore, the Landau theory of the quantization of the cyclotron orbits
of charged particles in magnetic fields cannot be used to explain the
quantum Hall effects, since the interactions between the mobile electrons
and the ion lattice had been ignored totally in the theory. As we have
discussed above, these interactions play an essential role in the Hall
effects.
\begin{figure}[tp]
\begin{center}
\resizebox{1\columnwidth}{!}{
\includegraphics{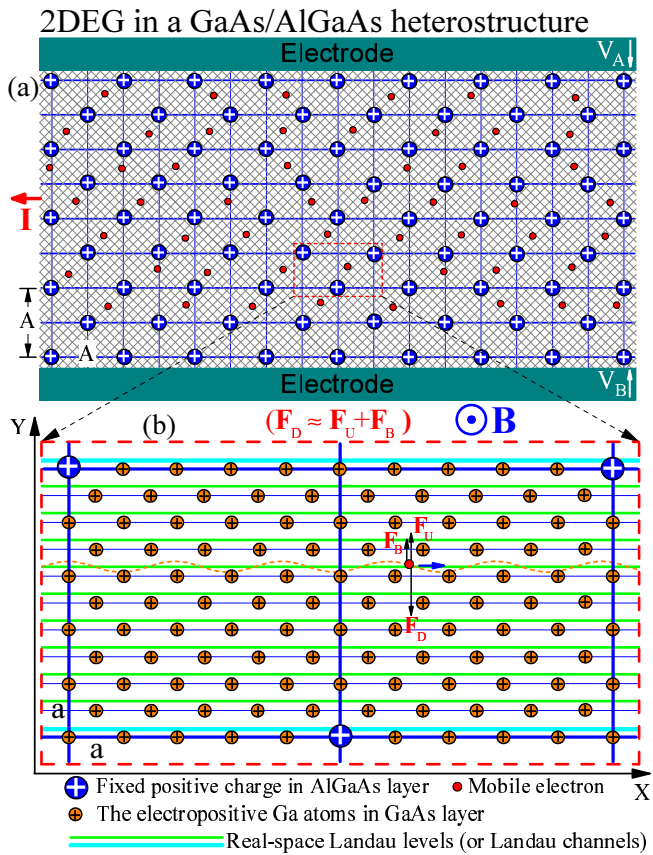}}
\end{center}
\caption{Two kinds of confinement interactions in a GaAs/GaAlAs
heterojunction. (a) The interaction between the mobile electrons and
positive ions inside GaAlas layer, (b) the local enlarged figure of (a), the
confinement of the electrons by the electropositive Ga atoms in GaAs layer
is shown and lattice confinement induced real space Landau channels are
suggested. }
\label{fig8}
\end{figure}

Here, we will try to provide a unified explanation of the IQHE and FQHE
based on our hypothesis. In a GaAs/GaAlAs heterojunction as shown in Fig. %
\ref{fig6}, the mobile electrons (or electron gas) is effectively restricted
within the heterojunction of $XY$-plane but their movement in $XY$-plane is
assumed to be free (the interactions between the mobile electrons and ion
lattice are neglected) in all the existing theories of Hall effect. One most
important feature of our theory is that without the interactions between the
mobile electrons and ion lattice, there is no Hall effect in the system. In
other words, to study the Hall effect we must take into account the
interactions between the mobile electrons and ion lattice.

Figure \ref{fig8} shows two kinds of interactions between the mobile
electrons and ion lattice in a GaAs/GaAlAs heterojunction. According to the
values of $n_{s}\sim 1.0\times 10^{11}/cm^{2}$ and the lattice constant of
GaAs $a=5.56\mathring{A}$, there is only one electron per unit supercell ($%
A\times A\approx 18a\times 18a$) of the superlattice, as shown in Fig. \ref%
{fig8}(a). Therefore, the mobile electrons will interact with the fixed
positive charges in AlGaAs layer. At the same time, these electrons are
confined by the electropositive Ga atoms in GaAs layer, see the enlarged
figure of Fig. \ref{fig8}(b). For a given external magnetic field, some real
space ballistic trajectories (or real space Landau levels) can be formed
along $X$-direction (the green and cyan solid lines), where the Lorentz
force $F_{B}$ can be well balanced by the electromagnetic interactions ($%
F_{D}=F_{U}+F_{B}$), as indicated in the figure. The combined effects of the
magnetic field and the positive ion lattice on the mobile electrons lead to
the \textquotedblleft localization\textquotedblright\ of the electrons in $Y$%
-direction with a real space real-space separation of approximately $a/2$
while extended in $X$-direction. But for now the most interesting question
is: How can such a picture explain the formation of Hall resistance plateaux
of Fig. \ref{fig7}? In our theoretical framework, the Hall resistance
satisfies
\begin{equation}
R_{xx}=\frac{V_{H}(B)}{I}\propto \xi (B).  \label{re1}
\end{equation}

As the electric current $I$ remains unchanged in the experiment, the
emergence of the Hall resistance plateaux means that the corresponding Hall
voltage $V_{H}$ does not change with magnetic field $B$. From Eq. (\ref{re1}%
), this imply that the applied magnetic field cannot induce a significant
displacement of the mobile electrons in some special positions along $Y$%
-direction. This conclusion is in agreement with the discussion above, when
the mobile electrons are moving along the restricted real space Landau
channels, they are very difficult to be shifted by the external magnetic
field ($\Delta \xi (B)\rightarrow 0$) and the Hall resistance plateaux
emerge. Furthermore, a wide plateau reveals a more intense confinement on
the electrons, if all the mobile electrons are traveling along the most
confined channels [indicated by cyan solid lines in Fig. \ref{fig8}(b)], a
wide plateau of the Hall resistance will naturally appear.

The final secret of the Hall effect: Why the Hall resistance plateaux take
on the quantized values related to the Planck's constant $h=6.626196\times
10^{-34}(J\cdot s)$? For the quasi-two-dimensional Hall system, Eq. (\ref{re}%
) of the Hall resistance can be reexpressed as

\begin{equation}
R_{xx}=\frac{V_{H}}{I}=-\kappa \frac{B}{n_{s}^{1-\alpha }eb^{2}d}=-\frac{%
\kappa eB}{n_{s}^{1-\alpha }eb^{2}d}\frac{1}{e^{2}},  \label{re2}
\end{equation}%
where $n_{s}$ is the surface mobile electron density.

Because the Planck's constant $h$ is an infinitely small amount, from Eq. (%
\ref{re2}), with the increasing of the magnetic field there is always
possible that the magnetic field $B_{1}$ makes the following formula true
\begin{equation}
\frac{\kappa eB_{1}}{n_{s}^{1-\alpha }eb^{2}d}=h.  \label{re3}
\end{equation}

Inserting Eq. (\ref{re3}) into Eq. (\ref{re2}), one gets the Hall resistance
plateau $R_{xx}(B_{1})=h/e^{2}$. When the magnetic field $B>B1$, one can
obtain a set of Hall resistance plateaux which are given by%
\begin{equation*}
R_{xx}(B)=\frac{h}{\eta e^{2}},
\end{equation*}%
where $\eta <1$ is a fractional number. When the applied magnetic field $%
B<B1 $, another set of Hall resistance plateaux can be described as

\begin{equation*}
R_{xx}(B)=\frac{h}{\zeta e^{2}},
\end{equation*}%
where $\zeta >1$ is an integer or a fractional number. It should be pointed
out that these Hall resistance plateaux are closely related to the Landau
channels of Fig. \ref{fig8}(b). The integer $\zeta $ corresponds to the
integer quantum Hall effect, while the fractional $\eta $ and $\zeta $
contribute to the fractional quantum Hall effect.

\section{ Conclusion}

In this paper, we have developed a new theory for describing Hall effect in
various electronic systems based on a pure electron picture. It has been
shown solidly that the interaction of the charge carriers and the ion
lattice play an essential role in the Hall effects. Our theory provides an
explicit explanation of the sign different of the Hall coefficient in the $N$%
-type and $P$-type semiconductors. We have attempted to uncover the physical
nature of the sign reversal Hall phenomena induced by both temperature and
magnetic field in different materials. Furthermore, we have considered that
the integer and fractional quantum Hall effects should share exactly the
same physical mechanism and a unified and simple picture of these two
effects has been provided. We are confident that the research may shed light
on the fundamental of the Hall effect.

\end{document}